\documentclass[prl,twocolumn,showpacs,superscriptaddress,reprint]{revtex4-1}

\usepackage{amsmath, amsfonts, amssymb}
\usepackage{enumerate}
\usepackage{graphicx}
\usepackage{color}

\newcommand{\up}{\uparrow}
\newcommand{\dw}{\downarrow}

\newcommand{\mean}[1]{\langle #1 \rangle}

\newcommand{\bB}{\mathbf{B}}
\newcommand{\bS}{\mathbf{S}}
\newcommand{\ba}{\mathbf{a}}
\newcommand{\bb}{\mathbf{b}}
\newcommand{\ket}[1]{\left|#1\right\rangle}
\newcommand{\bra}[1]{\left\langle#1\right|}

\begin{document}


\title{Entanglement detection from conductance measurements in carbon nanotube Cooper pair splitters}

\newcommand{\madrid}{Departamento de F\'{\i}sica Te\'{o}rica de la Materia Condensada,
Condensed Matter Physics Center (IFIMAC),
and Instituto Nicol\'{a}s Cabrera,
Universidad Aut\'{o}noma de Madrid, E-28049 Madrid, Spain}
\newcommand{\wurzburg}{Institute for Theoretical Physics and Astrophysics,
University of W\"{u}rzburg, D-97074 W\"{u}rzburg, Germany}

\author{Bernd Braunecker}
\affiliation{\madrid}

\author{Pablo Burset}
\affiliation{\madrid}
\affiliation{\wurzburg}

\author{Alfredo Levy Yeyati}
\affiliation{\madrid}

\date{\today}


\begin{abstract}
Spin-orbit interaction provides a spin filtering effect in carbon nanotube based Cooper pair
splitters that allows us to determine spin correlators directly from current measurements.
The spin filtering axes are tunable by a global external magnetic field.
By a bending of the nanotube the filtering axes on both sides of the Cooper pair splitter
become sufficiently different that a test of entanglement of the injected Cooper pairs
through a Bell-like inequality can be implemented.
This implementation does not require noise measurements, supports imperfect
splitting efficiency and disorder, and does not demand a full knowledge of the spin-orbit strength.
Using a microscopic calculation we demonstrate that entanglement detection by
violation of the Bell-like inequality is within the reach of current experimental setups.
\end{abstract}




\maketitle


The controlled generation and detection of entanglement is a necessary step toward the
goal of using quantum states for applications.
In a solid state nanostructure this control ideally allows us to
manipulate and detect entanglement between selected pairs of electrons.
A promising source of entangled electron pairs is the Cooper pair splitter (CPS).
It consists of a superconductor that injects Cooper pairs through two quantum dots (QDs)
into two outgoing normal leads, designed such that the Cooper pair electrons
preferably split and leave the superconductor over different
leads but preserve their spin entanglement \cite{recher:2001,cps}.
Very recently several CPS experiments have been performed
\cite{hofstetter:2009,herrmann:2010,hofstetter:2011,das:2012,schindele:2012} and
Cooper pair splitting efficiencies up to 90\% have been reached \cite{schindele:2012}.
So far, however, a proof that the electrons remain entangled is still lacking.

The present experiments do not allow to resolve individual splitting events, and
the results of the measurements are time averaged quantities, such as
current or noise. These provide information on the average
spin correlations of the injected Cooper pairs. In this Letter we demonstrate that
this information can be extracted from the currents alone
in a carbon nanotube (CNT) based CPS, if spin-orbit interaction (SOI) effects are
taken into account~\cite{cottet:2012}. This allows us to propose a general entanglement
test, based on the Bell inequality \cite{clauser:1969,bell_solid_state},
which does not require noise measurements \cite{note}.

Indeed, the SOI in CNTs leads to unique spin-energy filtering properties
that directly modulate the Cooper pair splitting current flowing out of the CPS,
and ideally suppress any noise.
From conductance measurements it is then already possible to reconstruct all
spin correlators contained in the Bell inequality, thus avoiding the need of
ferromagnetic contacts as spin filters, which are challenging to implement.
Without noise measurements we also
avoid the associated problem of electron fluctuations in the detectors~\cite{hannes:2008}.
The built-in energy filtering furthermore leads to an enhanced Cooper pair splitting
efficiency \cite{veldhorst:2010}.

\begin{figure}[b]
	\includegraphics[width=\columnwidth]{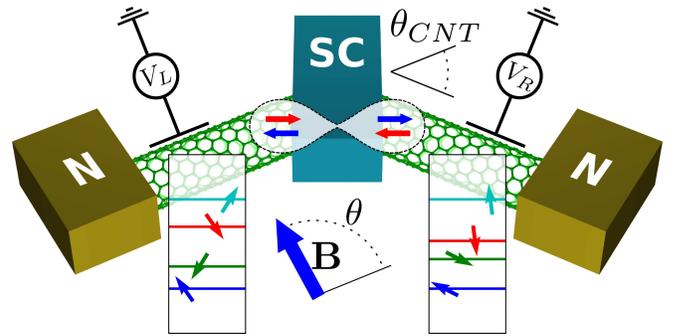}
	\caption{\label{fig:setup}
	Double quantum dot CPS based on a bent CNT in an external magnetic field $\bB$.
	Because of $\bB$, SOI, and the bending angle $\theta_{CNT}$ of the CNT, the
	spin-valley degeneracy of the QD levels is lifted, and the resulting 4 levels
	(boxes) are spin polarized as indicated by the arrows
	(see also Fig. \ref{fig:optimal}).
	The superconductor SC injects Cooper pairs (hourglass shape) that split onto
	the QDs and provide a current to the normal leads N that is modulated by the
	spin projections of the QDs (tunable by the gates $V_{L,R}$) and can be used
	to determine the spin correlators for the Bell inequality.
	}
\end{figure}

The proposed CPS setup is shown in Fig. \ref{fig:setup} and consists of
a regular double-QD CPS built from a single-wall CNT, yet made with a (naturally)
bent CNT such that there is an angle $\theta_{CNT}$ between the QD axes.
Alternatively, the QDs can be built from separate CNTs with similar diameters and
an angle $\theta_{CNT}$ between them.
The SOI spin splits the QD levels. In combination with a
global magnetic field $\bB$, the fourfold spin-valley degeneracy of the QD levels
is completely lifted. The split levels provide a unique spin filter for
electron transport with two spin projection axes per QD, filtering directly
the injected Cooper pair current. Therefore, conductance measurements alone,
at fixed $\bB$,
allow a reconstruction of all the spin correlators necessary for the Bell inequality.
The spin projection axes are different in the two QDs
due to the bending, and are tunable by $\bB$. In the following we show that this tunability
provides sufficient conditions for obtaining violations of the Bell inequality in an ideal CPS.
We then proceed to a full microscopic calculation and demonstrate
that the result remains robust under realistic conditions,
as achievable by present experiments.


\emph{SOI in CNT quantum dots}.
CNTs are graphene sheets rolled into a cylinder. They preserve the graphene
band structure with two Dirac valleys but have enhanced SOI contributions
due to the curvature.
The corresponding model, including the effect of $\bB$,
is described by the sum of the Hamiltonians \cite{izumida:2009,jeong:2009,klinovaja:2011a,klinovaja:2011b}
\begin{align}
	H_0     &= \hbar v_F \bigl[ k_t^0 \sigma_1 + k \tau_3 \sigma_2 \bigr],
\\
	H_{cv}  &= \hbar v_F \bigl[ \Delta k_t^{cv} \sigma_1 + \Delta k_z^{cv} \tau_3 \sigma_2 \bigr],
\\
	H_{SOI} &= \alpha \sigma_1 S_z + \beta \tau_3 S_z,
	\label{eq:SOI}
\\
	H_B     &= \mu_B g \bB \cdot \bS /2 + |e| v_F R B_z \sigma_1 /2,
	\label{eq:H_B}
\end{align}
which are matrices in the space spanned by the graphene sublattice indices
$\sigma=A,B$ (with Pauli matrices $\sigma_{1,2,3}$),
the valleys $\tau=K, K' = +, -$ (Pauli matrices $\tau_{1,2,3}$),
and the spin projections $S=\up,\dw$ (Pauli matrices $S_{x,y,z}$,
with $S_z$ oriented along the CNT axis).
$v_F$ is the Fermi velocity, $k_t^0$ the transverse quantized momentum (zero for
metallic CNTs), $k$ the longitudinal momentum,
$\Delta k_{t,z}^{cv}$ are momentum corrections
induced by the curvature,
$\alpha, \beta$ determine the SOI, $\mu_B$ is the Bohr
magneton, $g=2$ the Land\'e $g$-factor, $e$ the electron charge, $R$ the CNT radius,
and $B_z$ the component of $\bB$ along $S_z$.
We have neglected terms leading to the formation of Landau levels since
at the considered sub-Tesla fields they are of no consequence.
For a QD, $k$ is further quantized by the QD length \cite{bulaev:2008,weiss:2010,lim:2011}.

Because of its momentum independence, the SOI takes the role of an internal
valley (and QD orbital) dependent Zeeman field $\tau \bB_{SOI}$
along $S_z$,
which combines with $\bB$ to the effective field in each valley
$\bB^\tau_{\text{eff}} = \bB + \tau \bB_{SOI}$. These fields lift
the spin degeneracy of the QD levels, while the orbital effect of Eq. \eqref{eq:H_B}
lifts the energy degeneracy between the two valleys for any $B_z \neq 0$.
The QD levels turn into spin-valley-energy filters.
The effective fields define the spin polarization axes
$\ba_\tau \propto \bB_{\text{eff}}^{\tau}$,
which are nonparallel if $\bB^{K}_{\text{eff}} \neq \bB^{K'}_{\text{eff}}$,
tunable by $\bB$,
and such that the spin-eigenstates in each valley $\ket{\pm a_\tau}$
fulfill $(\bS\cdot\ba_\tau)\ket{\pm a_\tau} = \pm \ket{\pm a_\tau}$
(full polarization).
If $P_{\pm a_\tau} = \ket{\pm a_\tau} \bra{\pm a_\tau}$, spin measurements
can be reconstructed by electron transport over the different QD
levels by $(\bS\cdot\ba_\tau) = P_{+a_\tau}-P_{-a_\tau}$.


\begin{figure}
	\includegraphics[width=\columnwidth]{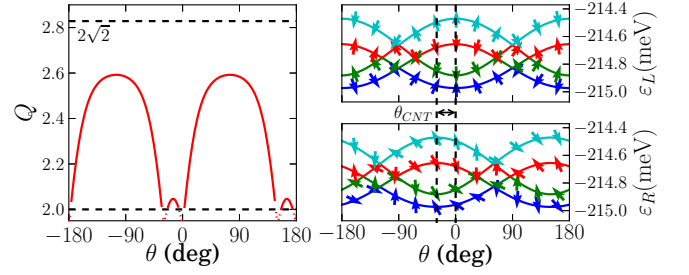}
	\caption{\label{fig:optimal} 
	Values $Q$ of the Bell equation \eqref{eq:Q} (left panel) for an ideal bent CNT-CPS as a function
	of in-plane $\bB$-field rotation angle $\theta$,
	for the lowest valence band orbitals in a CNT of chirality (18,10),
	$|\bB| = 0.4$ T, $\theta_{CNT}= 30^\circ$,
	$\alpha=-0.08$ meV, $\beta = -0.15$ meV,
	and QD lengths of 200 nm.
	For this situation, $|\bB|/|\bB_{SOI}| \approx \mu_B g |\bB|/ 2 |\alpha-\beta|=0.34$.
	The horizontal lines mark the threshold $Q=2$ and the maximal possible $Q=2\sqrt{2}$.
	The right panels show the $\theta$ dependence of the level energies of both QDs.
	The spectra are identical up to the shift by $\theta_{CNT}$ marked by the vertical dashed lines.
	The arrows indicate the spin polarizations
	in a global spin basis, as used for the determination of $Q$.
	}
\end{figure}


\emph{Bell test in an ideal CNT-CPS}.
In the double-QD system shown in Fig. \ref{fig:setup}, the CNT bending
changes the orientation of $\bB_{SOI}$ and so of $\bB_\text{eff}^\tau$.
The spin polarization axes $\ba_\tau$ in the left QD become distinct from the axes
in the right QD, which we call $\bb_\tau$.
We consider an ideal CPS, characterized by a perfect
Cooper pair splitting efficiency with valley-independent pair injection (see discussion
below) and isolated sharp QD levels.
Since any injected Cooper pair splits onto the different levels in each QD
(the current consists only of split Cooper pairs),
and the tunneling amplitude onto each dot is proportional to the spin projection,
the current collected at the normal leads in resonant conditions for a given pair
of levels is proportional to
$\mean{P_{\pm a_\tau} \otimes P_{\pm b_\tau'}}$ and allows us to
reconstruct the spin correlators
$C_{\ba_\tau,\bb_{\tau'}} =
\mean{(\bS\cdot\ba_\tau)\otimes(\bS\cdot\bb_{\tau'})}$
[see Eq. \eqref{eq:C}].
The availability of 2 spin projection axes per QD consequently allows us
to test the CHSH-Bell inequality \cite{clauser:1969}
\begin{equation} \label{eq:Q}
	Q = | C_{\ba_{K},\bb_{K}} + C_{\ba_{K},\bb_{K'}} + C_{\ba_{K'},\bb_{K}} - C_{\ba_{K'},\bb_{K'}} |
	\le 2.
\end{equation}
Any non-entangled state (including the steady state density matrix considered here) 
fulfills this inequality. A violation
$Q>2$ is
sufficient
to prove entanglement.
For a spin-singlet, a maximal $Q=2\sqrt{2}$ is obtained by orthogonal
$\ba_K \perp \ba_{K'}$, $\bb_K \perp \bb_{K'}$, and $45^{\circ}$ between $\ba_K$
and $\bb_K$.
Such optimal axes cannot be generally obtained in the CNT-CPS, for which $\bB_{SOI}$ and
$\theta_{CNT}$ are fixed by the sample fabrication, and only $\bB$ is tunable.
Yet, as we show in Fig. \ref{fig:optimal}, this tunability is sufficient to obtain
$Q>2$ as a function of the angle $\theta$ of a rotating
in-plane field $\bB = B (\sin\theta,0,\cos\theta)$  (see Fig. \ref{fig:setup}),
for $B \sim |\bB_{SOI}|$.
The shown result is generic and we find similar $Q>2$ for most CNT chiralities,
diameters, and QD lengths.


\emph{Realistic systems}.
In a realistic setup, the two QDs remain coupled
through the superconducting region, their levels are broadened by the contacts,
the splitting efficiency is imperfect and electron pairs
can tunnel onto the same QD, the tunneling rates depend on the gate voltages, and electrons
can interact.
Any measurement probes the steady state density matrix $\rho$ of the full CPS system and not an
ideal singlet state. The projections $P_{\pm a_\tau}, P_{\pm b_{\tau'}}$
are obtained by narrowing the measurement to an energy window 
capturing
the electron transport through the corresponding level of each QD, typically
by differential conductance measurements tuned to the resonances corresponding to the levels.
The modified $\rho$ together with the measurement method leads to a distorted
reconstruction of the spin correlators, and we need to distinguish between \emph{local}
and \emph{nonlocal} distortion sources.

Local distortions in one QD are independent of the other QD
and modify, e.g., $P_{\pm a_\tau}$ to $P_{+ a_\tau'}, P_{-a_\tau''}$.
We can write
$P_{+a'_\tau}-P_{-a''_\tau} = \gamma (\bS\cdot\tilde{\ba}_\tau) + (1-\gamma) P_{\tilde{\ba}_\tau}$
for an intermediate axis $\tilde{\ba}_\tau$, $0 \le \gamma \le 1$, and a
remaining projection $P_{\tilde{\ba}_\tau}$. The latter transforms any state into
a product state, and local distortions therefore lower the
ideal value of $Q$ by an amount set by the various $\gamma$ for the different QD levels.
Assuming that the level broadening can be kept small so that there is only little overlap
between nearby resonances (assisted also by a charging energy),
the most important source of local distortions is disorder scattering within
each QD. It mixes the wave functions in different valleys \cite{kuemmeth:2008,jespersen:2011},
and the $\ket{\pm a_\tau}$ are no longer the eigenstates.
While of central importance in metallic CNTs, in semiconducting CNTs
disorder scattering competes with the valley-preserving semiconducting gap
of typically $\sim$ 100 meV,
which has opposite signs in opposite valleys.
If the disorder scattering amplitude is smaller
it has a negligible influence.
Therefore, semiconducting CNTs are preferable for testing the Bell inequality.

Valley mixing at injection, however, is essential. Indeed, if valleys and spins
are correlated, for instance, if the singlet splits
always into opposite valleys, the transport through other valley
combinations does not provide any information on the Cooper pairs and
the construction of $Q$ is no longer possible. For
a valid spin correlator measurement the injection must mix valleys to produce
a detectable signal through all resonances, yet the precise degree of mixing is unimportant.

Nonlocal distortions of the spin modify the spin projections
as an effect of the entire CPS system, typically by hybridization between the two QDs,
and the measured $P_{\pm a_\tau}, P_{\pm b_{\tau'}}$ become nonlocal operators.
Such operators can generate additional entanglement through wave function
mixing between the left and right QDs. In the CPS setup they
are a source of error for detecting spin entanglement.
Yet with the full microscopic calculation discussed next
we can see that these nonlocal contributions can be kept under control
in realistic conditions.


\begin{figure}
	\includegraphics[width=\columnwidth]{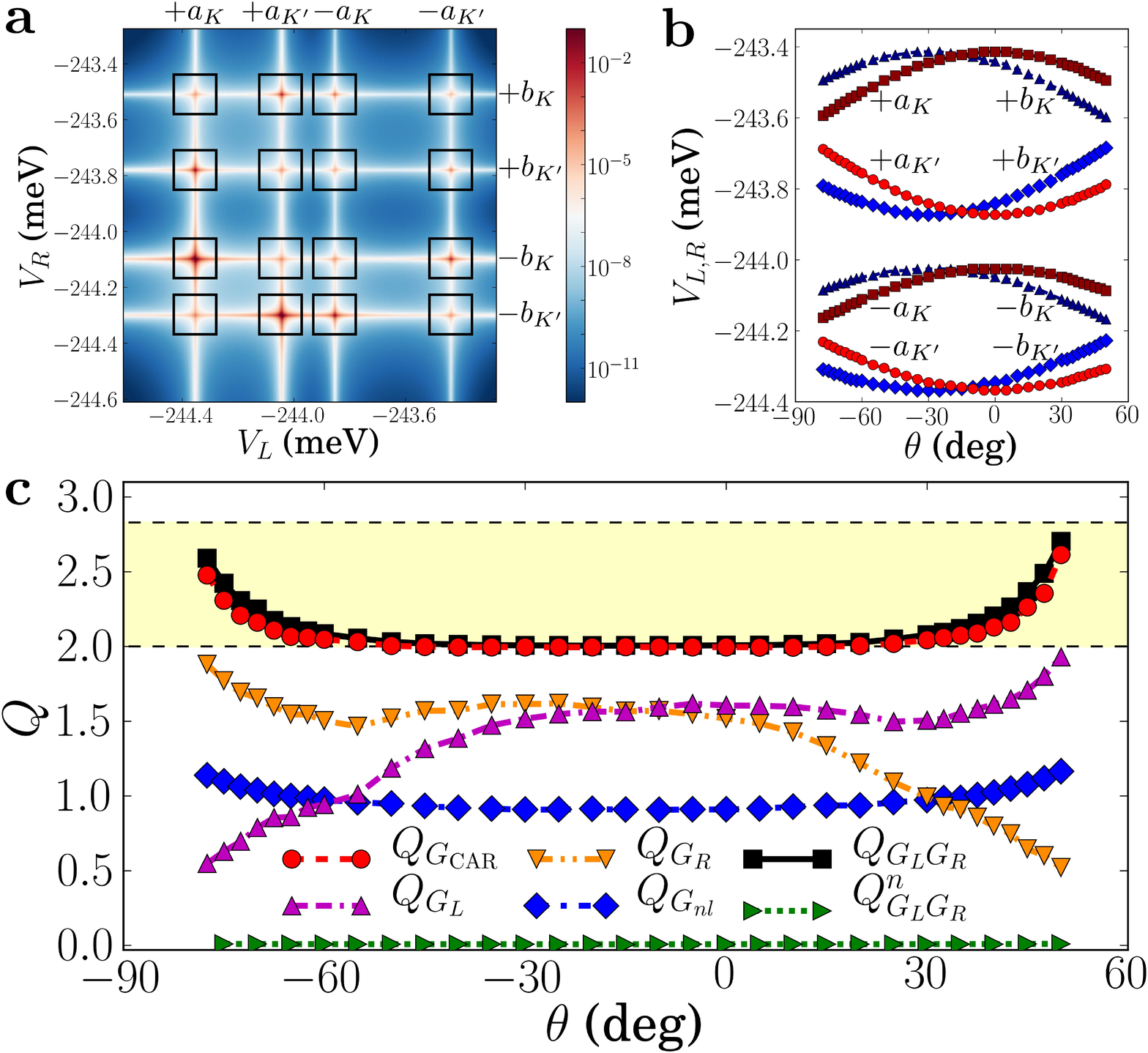}
	\caption{\label{fig:results}
	Results from the microscopic calculation of a CPS, based on a zigzag CNT of chirality
	(20,0) with a bending angle $\theta_{CNT}\approx 30^\circ$, in a field of $|\bB| = 0.5$ T
	(see the Supplemental Material \cite{supplement}).
	The SOI energies $\alpha = -0.10$ meV and $\beta = - 0.40$ meV
	lead to $|\bB|/|\bB_{SOI}| \approx \mu_B g |\bB|/ 2 |\alpha-\beta| = 0.10$.
	(a) Map of the conductance product $G_L G_R$ (units of $e^4/h^2$)
	as a function of QD gate voltages $V_{L,R}$ at $\theta =25^\circ$. The
	4 levels of each QD give rise to the 4 resonances labeled by $\pm a_\tau$, $\pm b_{\tau'}$.
	Inside the black squares, the CPS acts as a spin-valley filter for
	the projections $P_{\pm a_\tau} P_{\pm b_{\tau'}}$, and integrating the signal
	within each black square yields the corresponding observable.
	(b) $V_{L,R}$ values marking the positions of the resonances of the levels $\pm a_\tau$
	and $\pm b_{\tau'}$ of the two QDs as a function of $\theta$. The curves are identical
	up to the shift by $\theta_{CNT}$.
	Levels in the same valley $\tau$ see the same field $\bB^\tau_\text{eff}$ and are
	identified by having the same curvature as function of $\theta$.
	(c) $Q$ as a function of $\theta$
	for the conductances $G$ given as subscripts of $Q$ in the figure legend.
	The $Q$ values are obtained by analyzing data as shown in panel (a)
	by the method described in the text.
	The yellow shaded region
	marks the allowed range of violation of the Bell inequality for the spin-singlets
	in the steady state.
	}
\end{figure}


\emph{Microscopic model}.
To quantitatively access a realistic system and to determine the
optimal choice of measurements that allows us to gain insight in the effects of
local and nonlocal distortions, we have investigated
a microscopic tight-binding model of the CNT-CPS. Our approach follows Ref.~\onlinecite{burset:2011},
which we have complemented to include magnetic fields by terms equivalent to Eq. \eqref{eq:H_B}
and valley mixing at injection.
As a result, we obtain the partial conductances of the CPS due to
Cooper pair splitting (crossed Andreev reflections, $G_\text{CAR}$), elastic cotunneling
through the superconducting region ($G_\text{EC}$), and the local Andreev scattering contributions
at each QD ($G_{\text{A}L},G_{\text{A}R}$).
From these quantities, transport from the superconductor to the normal leads is expressed by
the conductances $G_j = 2(G_{\text{A}j} + G_\text{CAR})$ ($j=L,R$), and transport between
the normal leads by the nonlocal conductance
$G_{nl} = G_\text{EC}- G_\text{CAR}$.

In Fig. \ref{fig:results} (a) we display a conductance map for a semiconducting CNT
as a function of the QD gate voltages $V_{L,R}$ that tune the QD levels to resonance.
Such a result is useful for a Bell test if all 4 resonances in each QD are well resolved
and their 16 points of intersection, corresponding to the products
$P_{\pm a_\tau} \otimes P_{\pm b_{\tau'}}$, form single peaks and not
avoided crossings. To access this regime, we have chosen a coupling between
the superconductor and the CNT on the order of the superconducting gap ($\lesssim$ 1 meV),
and tuned the coupling to the leads such that the resonances are well resolved (see the 
Supplemental Material \cite{supplement}).
Similar conditions have been obtained in experiments \cite{kuemmeth:2008,jespersen:2011}, and
such a regime can be reached for a wide variety of samples and coupling strengths to the
contacts.

To analyze the data we integrate the various conductances over regions centered at
the crossings as shown by the black squares in Fig. \ref{fig:results} (a).
From the resulting 16 integrals $G_{\pm a_\tau,\pm b_{\tau'}}$
we construct the spin correlators
\begin{equation} \label{eq:C}
C_{\ba_\tau,\bb_{\tau'}} =
\frac{\sum_{\nu,\nu'=\pm} \nu\nu' G_{\nu a_\tau,\nu' b_{\tau'}}}
{\sum_{\nu,\nu'=\pm}G_{\nu a_\tau,\nu' b_{\tau'}}},
\end{equation}
which is a simple consequence from the fact that $P_{+a_\tau}-P_{-a_\tau} = (\bS\cdot\ba_\tau)$
and $P_{+a_\tau}+P_{-a_\tau}$ is the identity operator (see the Supplemental Material \cite{supplement}).
From these $C_{\ba_\tau,\bb_{\tau'}}$
we determine $Q$ by Eq. \eqref{eq:Q}, with the liberty of
placing the $-$ sign in front of any term in Eq. \eqref{eq:Q} to obtain the
maximum $Q$.

The Cooper pair splitting amplitude is directly described by $G_\text{CAR}$,
and the corresponding curve $Q_{G_\text{CAR}}$ [Fig. \ref{fig:results} (c)] captures
indeed a similar behavior as the ideal case of Fig. \ref{fig:optimal},
with $Q>2$ in the $\theta$ regions where the levels of different valleys approach each other and
the spin projections rotate [Fig. \ref{fig:results} (b)].
The measurable conductances $G_j$, however, contain with $G_{\text{A}j}$ contributions
that represent strong enough local distortions to suppress $Q$ below 2.
In the right QD the local distortions are enhanced by level overlaps close to
$\theta = 60^{\circ}$ where the $K$ and $K'$ levels become degenerate
[Fig. \ref{fig:results} (b)], and indeed $Q_{G_R}$ decreases
in this region. In contrast, the left QD levels remain well separated, and
$Q_{G_L}$  mirrors the upturn of $Q_{G_\text{CAR}}$, with $G_\text{CAR}$ overruling the
$G_{\text{A}L}$ contribution. The same behavior with $G_L \leftrightarrow G_R$ is found
near $\theta = -90^\circ$.
On the other hand,
$G_{nl}$ corresponds to an experiment of electron injection through a normal lead
and contains with $G_\text{EC}$ a component describing the uncorrelated single-particle
transport. Since we find that $G_\text{EC}$ and $G_\text{CAR}$ have a similar amplitude,
we expect that $Q_{G_{nl}} \sim Q_{G_\text{CAR}}/2$.
However, $G_\text{EC}$ contains also the higher order tunneling processes
that represent the nonlocal distortions, which may cause $Q_{G_{nl}}$ to increase again.
Nonetheless, we find that $Q_{G_{nl}} \sim 1$ with a similar shape as $Q_{G_\text{CAR}}$,
indicating that the nonlocal distortions have a negligible effect.

While $G_\text{CAR}$ produces the purest indicator of spin entanglement, it is only indirectly
accessible by experiments. On the other hand, the directly measurable
$G_{j}$ are obscured by the local contributions
of the $G_{\text{A}j}$. A method of circumventing this problem is
to consider products of the $G_j$, such as $G_L G_R$. Since the projections
$P \equiv P_{\pm a_\tau} \otimes P_{\pm b_{\tau'}}$
eliminate all QD degrees of freedom, the product $G_L G_R$
is equivalent to a nonlocal current measurement with a density matrix
$\rho'$ whose nonlocal contribution is encoded in $P \rho' P \propto P \rho^2 P$.
By the higher power of $\rho$ and the projections, the relative weight of
the local contributions can be reduced, while a spin singlet in $P \rho P$
remains a spin singlet in $P \rho^2 P$. In Fig. \ref{fig:results} (c)
we see that the corresponding curve $Q_{G_L G_R}$ follows almost perfectly
$Q_{G_\text{CAR}}$, showing that the multiplication $G_L G_R$
is powerful enough to suppress the local distortions in the $G_j$.
Therefore, a high splitting efficiency of a CPS is not a primary
requirement for the proposed Bell test.

To demonstrate that the large $Q$ value is indeed an effect of superconductivity,
we show with $Q_{G_LG_R}^n$ the corresponding curve for $G_LG_R$ obtained for
the normal state. The fact that $Q_{G_LG_R}^n \approx 0$ is the strongest indicator
that $Q_{G_LG_R}$ demonstrates indeed the spin entanglement.

Finally, we have truncated the curves in Fig. \ref{fig:results}
close to $\theta = 60^\circ$ and $-90^\circ$ where QD levels strongly overlap
[Fig. \ref{fig:results} (b)] and spin correlators can no longer be reconstructed.
It is indeed important to maintain well separated QD levels. Hence the charging energy
of the QDs, which has been neglected in the microscopic calculation, plays here
an important role as it increases the level separation but has much reduced exchange
coupling due to the SOI induced spin projections of the QD levels.


\emph{Conclusions}.
We have demonstrated that due to SOI effects bent CNT-CPS (or two CNTs under an angle)
can be used for entanglement detection in the steady state by a violation of the Bell inequality.
Notable for the Bell inequality is that the set of axes $\ba_\tau,\bb_{\tau'}$ along
which the spin correlators must be measured can be arbitrary and the precise axis orientations, i.e.,
the precise SOI strengths,
do not need to be known. This is an advantage over entanglement witnesses \cite{faoro:2007} or
quantum state tomography.
Although discussed for CNTs, the introduced concept of entanglement detection is general
and can be implemented in any system allowing tunable spin-energy filtering.
For an ideal CNT-CPS, a violation of the Bell inequality can be achieved for most CNTs
over a large range of orientations of an external field $\bB$ with strength $B \sim |\bB_{SOI}|$,
which for usual CNTs are $< 1$ T.
The robustness of this behavior was confirmed by a microscopic calculation
that incorporates the local and nonlocal imperfections of a realistic system. From the
results we propose the use of the product of conductances $G_L G_R$ as the optimal observable
for testing the Bell inequality. We have furthermore argued that the spin reconstruction in
semiconducting CNTs is robust against disorder.

To conclude, it should be noted that a bending of the CNT is not an absolute requisite.
An equivalent effect can be obtained by applying individual $\bB$ fields on the
QDs or by providing a constant field offset on one QD by placing a ferromagnet in its
vicinity, if sufficient control of the typical field strengths $|\bB| \sim |\bB_{SOI}| < 1$ T
can be granted. If two separate CNTs are connected to the superconductor, they
should have similar diameters such that their $\bB_{SOI}$ are comparable.


\emph{Acknowledgments}.
We thank A. Baumgartner, J. C. Budich, A. Cottet, N. Korolkova, P. Recher, and B. Trauzettel
for helpful discussions and comments.
We acknowledge the support by the EU-FP7 project SE2ND [271554]
and by the Spanish MINECO through Grant No. FIS2011-26516.
P.B. also acknowledges the support by the ESF under the EUROCORES Programme EuroGRAPHENE.





\renewcommand*{\citenumfont}[1]{S#1}
\renewcommand*{\bibnumfmt}[1]{[S#1]}

\setcounter{equation}{0}
\setcounter{figure}{0}
\renewcommand{\thefigure}{S\arabic{figure}}

\onecolumngrid

\vspace*{1.0cm}

\begin{center}
{\Large\sc Supplemental Material}
\end{center}

\vspace*{0.75cm}

\twocolumngrid


\section{Demonstration of Eq. (6)}

The current flowing out of an ideal CPS originates only from split Cooper pairs, with one 
electron being transported over the left and one electron over the right QD. 
This current is, therefore, subjected to the filtering of spin, valley, and energy of \emph{both} QDs,
and probing the current \emph{locally} in one QD contains the \emph{nonlocal} information of the filtering 
effects of both QDs. 

Indeed, in this situation, with filters set along the axes $\nu a_\tau, \nu' b_{\tau'}$
($\nu,\nu' = \pm$) and resonant conditions such that transport is restricted to the selected levels, 
the density matrix for the outflowing particles takes the form
$\rho_{\nu a_\tau,\nu' b_{\tau'}} = P_{\nu a_\tau} P_{\nu' b_{\tau'}} \rho P_{\nu' b_{\tau'}} P_{\nu a_\tau}$, with
$\rho$ the density matrix in the absence of spin-valley filtering. Due to the perfect splitting efficiency, the currents 
through the left and right QD are identical, and we can focus, for instance, on transport through the left QD only.
If $\hat{I}_L$ is the spin and valley independent current operator for transport over the left QD, 
the property $[\hat{I}_L, P_{\nu a_\tau} P_{\nu' b_{\tau'}}]=0$ ensures that 
$\langle \hat{I}_L\rangle = \mathrm{Tr}\{ P_{\nu a_\tau} P_{\nu' b_{\tau'}} \hat{I}_L \rho P_{\nu' b_{\tau'}} P_{\nu a_\tau}\} 
= \mathrm{Tr}\{ P_{\nu a_\tau} P_{\nu' b_{\tau'}} \hat{I}_L \rho\}$. 
In the linear response regime we have furthermore $\langle \hat{I}_L \rangle = V G_L$, with $G_L$ the conductance and 
$V$ the voltage applied to both leads with respect to the superconductor. 
As a function of both QD gate voltages, $G_L$ is resonant at the level crossing
$\nu a_\tau, \nu' b_{\tau'}$. The full amplitude of the transport at this 
level crossing, denoted by $G_{\nu a_\tau, \nu' b_{\tau'}}$, is obtained by integrating $G_L$ over this resonance.
If furthermore the tunneling rates to the QDs are independent of the QD gates, the quantities
$\langle P_{\nu a_{\tau}} P_{\nu' b_{\tau'}} \rangle 
= G_{\nu a_\tau, \nu' b_{\tau'}} / \sum_{\tilde{\nu},\tilde{\nu}'} G_{\tilde{\nu} a_\tau, \tilde{\nu}' b_{\tau'}}$
allow us to reconstruct the spin correlators due to the identites
$(P_{+ a_\tau} - P_{- a_\tau}) \otimes (P_{+ b_{\tau'}} - P_{- b_{\tau'}}) 
= (\bS \cdot \ba_\tau) \otimes (\bS \cdot \bb_{\tau'})$ and
$(P_{+ a_\tau} + P_{- a_\tau}) \otimes (P_{+ b_{\tau'}} + P_{- b_{\tau'}}) 
= \openone \otimes \openone$. As a consequence we obtain Eq. (6) in the main text.
The relation between conductances and spin correlators, therefore, follows from the same considerations
used in the proposed entanglement tests based on noise measurements \cite{Scps,Sbell_solid_state}.

To further test Eq. (6) and its consequences on entanglement detection under realistic
conditions, we have implemented the microscopic numerical calculation.
As discussed in the main text, the numerical results give 
an objective demonstration that Eq. (6) and the conclusions
for entanglement detection remain robust.


\section{Influence of realistic setup on $Q$}

In this part of the supplement we illustrate the influence of the coupling of the CNT
to the superconductor and the normal leads on the determination of $Q$.
We provide all parameters used for the tight-binding calculation following
Ref. \onlinecite{Sburset:2011}.
Finally we show how the level energies and the
spin projections evolve with the magnetic field.

\begin{figure}
	\includegraphics[width=\columnwidth]{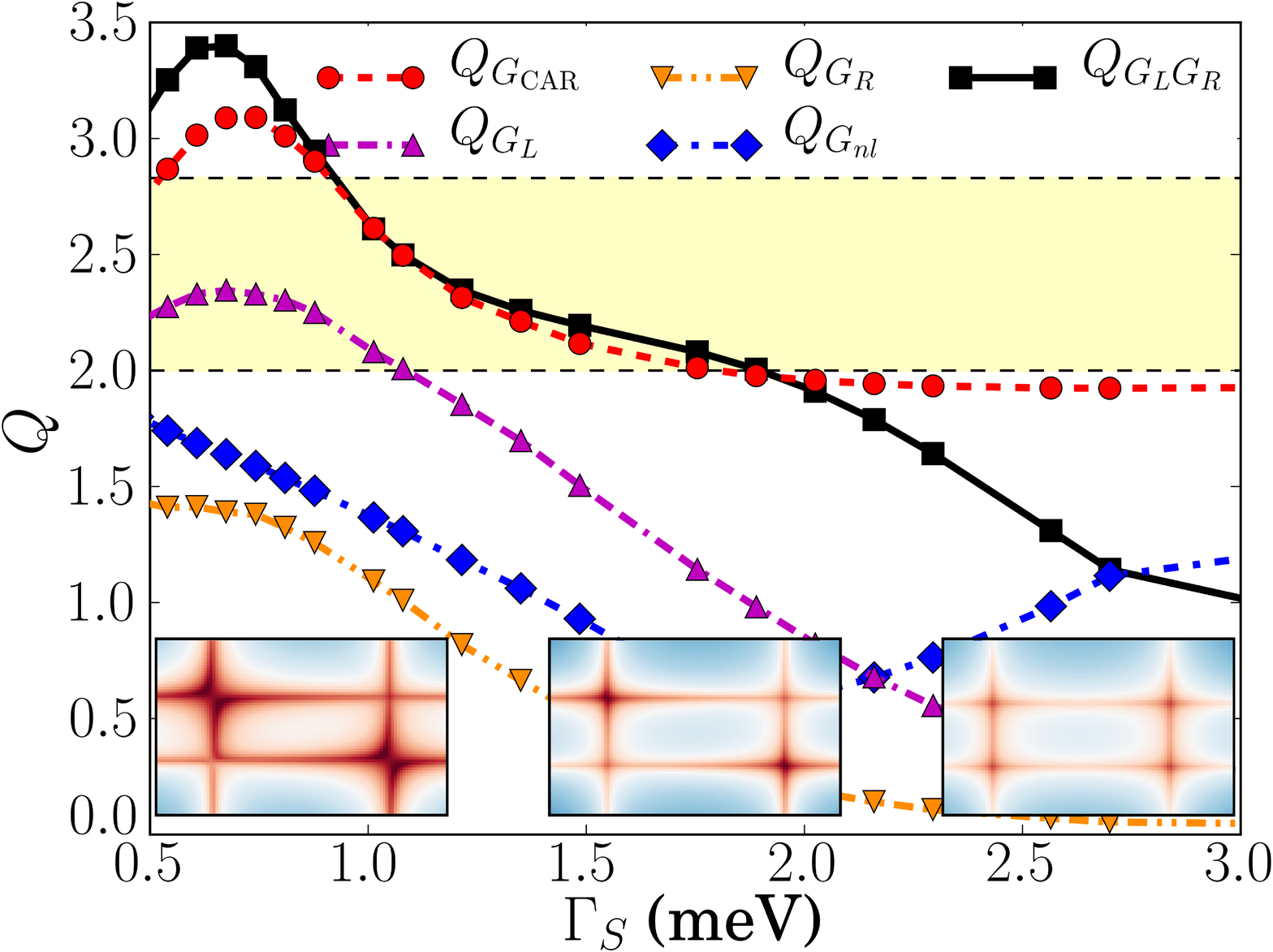}
	\caption{\label{fig:Gamma_S}
	Dependence of $Q$ on the effective coupling $\Gamma_S$ to the superconductor
	for a (20,0) CNT with fixed $\Gamma_{L,R} = 27$ meV, $B=0.5$ T, $\theta = 45^\circ$,
	$\theta_{CNT}=28.8^\circ$.
	The insets show a zoom on the conductance maps for the $\Gamma_S$ values corresponding
	to their placement in the figure, with identical logarithmic color scales [see Fig. 3 (a)
	in the main text].
	The center inset represents the valid regime
	for testing the Bell inequality with well resolved resonances of different
	intensities, and the absence of notable avoided crossings of the resonance peaks.
	}
\end{figure}

Figure \ref{fig:Gamma_S} shows the dependence
of $Q$ on the effective coupling strength $\Gamma_S$ between
the superconductor and the CNT. The insets show parts of the conductance
maps for the $\Gamma_S$ values corresponding roughly to the
placements of the insets in the plot. For large $\Gamma_S$, the
level broadening induced by the superconducting contact mixes the
Cooper pairs between the QD levels and the conductances are no
longer spin projective. This is notable by the similar intensities
of all resonances, and corresponds to a strong enhancement of the local
distortions discussed in the text. The corresponding values of $Q$
lie well below $2$. Small $\Gamma_S$, on the other hand, lead to a
weak Cooper pair injection amplitude compared with the hybridization
through the superconducting region. As a consequence, the resonance
crossings turn into anticrossings. The resulting $Q$ values sharply
increase beyond $Q=2\sqrt{2}$ due to strongly distorted
spin correlator reconstructions by the nonlocal hybridization processes.
At very small $\Gamma_S$, the anticrossings of different levels
overlap, and the spin correlator reconstruction becomes
erratic.

A valid measurement of $Q$ requires $\Gamma_S$ corresponding to
the central inset in the Fig. \ref{fig:Gamma_S}, represented by well-defined level
crossing peaks with unequal intensities. The unequal intensities are
a result from the spin filtering of the singlet states, such that
spin projection axes that are close to parallel suppress the conductance,
while projections that are close to antiparallel allow a maximal
transmission. Hence unequal, $\theta$ dependent peak intensities
are a necessary indicator for spin entanglement, and indeed are the
basis for the implementation of the Bell test.

\begin{figure}
	\includegraphics[width=\columnwidth]{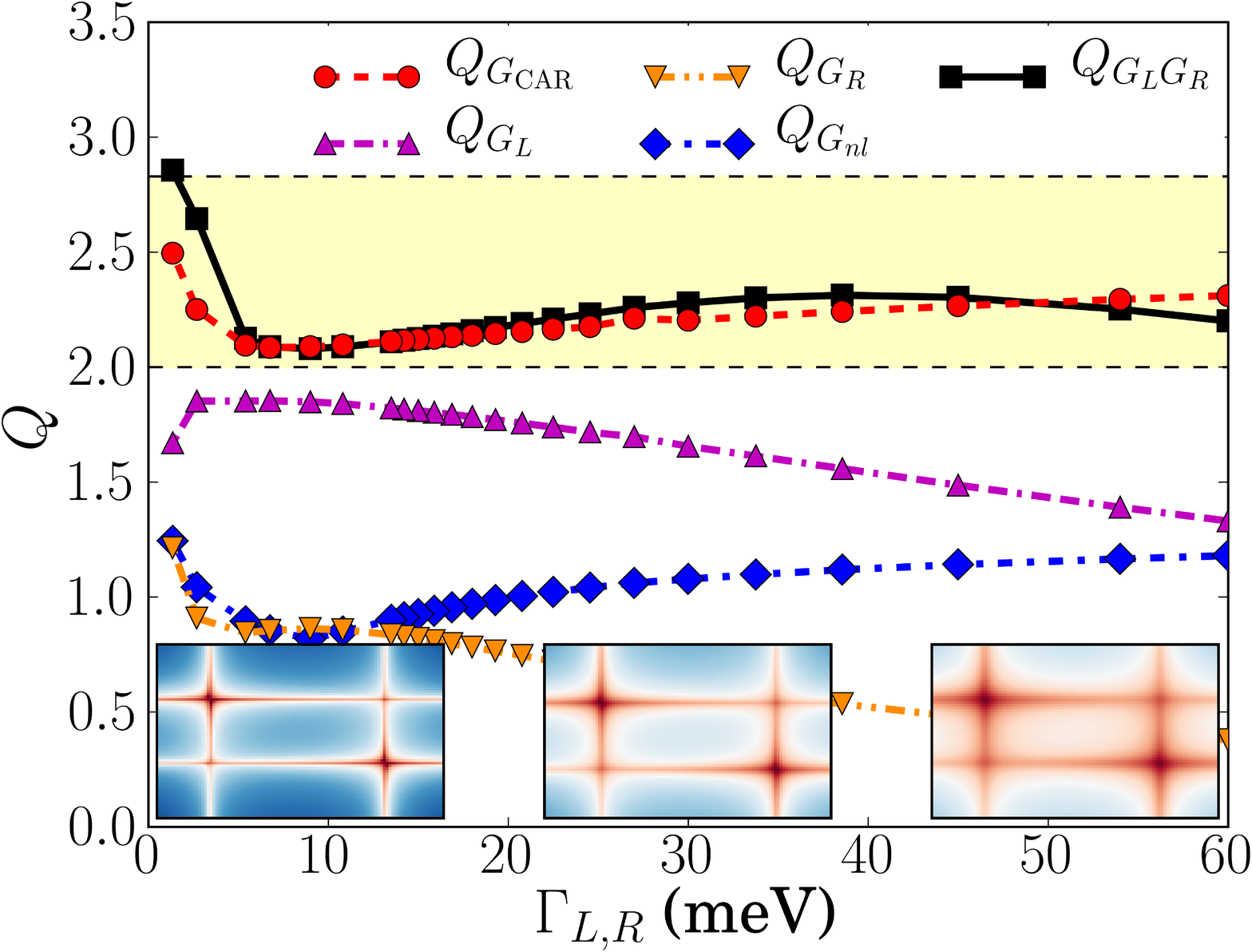}
	\caption{\label{fig:Gamma_LR}
	Dependence of $Q$ on the tunneling amplitudes $\Gamma_j$ to the normal lead $j=L,R$,
	for $\Gamma_L=\Gamma_R$
	for a (20,0) CNT with fixed $\Gamma_S = 1.35$ meV, $B=0.5$ T, $\theta = 45^\circ$,
	$\theta_{CNT}=28.8^\circ$.
	The insets show a zoom on the conductance maps for the $\Gamma_j$ values corresponding
	to their placement in the figure, with identical logarithmic color scales [see Fig. 3 (a)
	in the main text].
	The center inset represents the valid regime
	for testing the Bell inequality with well separated resonances and a high enough
	pixel resolution such that the integral weight of each peak can be determined with
	high accuracy.
	}
\end{figure}

The dependence on the tunnel coupling to the normal leads,
characterized by a tunneling amplitude $\Gamma_j$ for $j=L,R$, is
represented in Fig. \ref{fig:Gamma_LR}.
The combination of the $\Gamma_j$
with $\Gamma_S$ defines the broadening of the QD levels.
Indeed, in the model of Ref. \onlinecite{Sburset:2011} the lateral leads were
represented by ideal one-dimensional channels weakly coupled to each end site
of the nanotube. In the present calculations the tunneling rates to these leads
$\Gamma_{j}$ take values between 10 and 100 meV. The actual broadening introduced
to the QD levels becomes then on the order of $\Gamma_{j} a /W_{j}$ with $W_j$
the length of QD $j$ and $a$ the lattice constant.

In contrast to $\Gamma_S$, the insets in Fig. \ref{fig:Gamma_LR} show that $\Gamma_j$ contributes
only to a broadening of the levels but leaves the inequality of the
peaks unchanged. The $\Gamma_j$ values of the insets correspond again roughly
to the positions of the insets. At large $\Gamma_j$, the level overlaps
lead to strong local overlaps of the projections such that
the $Q_{G_j}$ strongly decrease. Since, however, the unequal
intensities and so the spin-filtering properties of each QD
level are maintained, the value of $Q_{\text{CAR}}$
remains large even for large $\Gamma_j$. Yet for larger $\Gamma_j$ the
influence of the overlaps is well notable by the split off of
$Q_{G_L G_R}$ from the $Q_{\text{CAR}}$ value.
For small $\Gamma_j$ we notice that most conductances lead to
an upturn of $Q$. This effect is attributable to the finite
resolution of the peaks from the numerics
that become only a few pixels wide, and the
result is strongly susceptible to the discretization steps of the $V_j$.
The artificial nature of the low $\Gamma_j$ behavior is indeed seen by the
comparison of $Q_{G_L}$ and $Q_{G_R}$, which show an anomalous opposite behavior in
a regime where all resonances are well separated and all couplings to the left
and right QDs are identical.
Finally, we notice that since the $\Gamma_j$ mainly influence the QD levels
locally, an asymmetry $\Gamma_L \neq \Gamma_R$ has only little impact on
the value of $Q$ as long as all levels can be well resolved.

The results shown in the main text represent the optimal values
for the chosen CNT and geometry, $\Gamma_S = 1.35$ meV and $\Gamma_L = \Gamma_R = 27$ meV,
determined by first identifying a valid $\Gamma_S$ leading to well shaped
peaks with modulated intensities, and then optimizing the $\Gamma_j$
to obtain well resolved resonances.
These values, however, are strongly sample and geometry dependent and can be used only as indicative.

For the present calculation we have used a CNT of chirality (20,0) with
QD lengths $W_L=W_R=43$ nm and a length of the central superconducting region
of 173 nm.
Yet the same behavior of level separations and $Q$ values is found for
longer system sizes corresponding to experimental situations.
A magnetic field of strength $B=0.5$ T was applied to each QD region
with angles $\theta$ on the left QD and angles $\theta+\theta_{CNT}$
on the right QD with respect to the CNT axis, for $\theta_{CNT} = 28.8^\circ$.
The SOI strengths $\alpha, \beta$ and the shift $\Delta k^{cv}_t$
have been implemented using the values of
Refs. \onlinecite{Sklinovaja:2011a,Sklinovaja:2011b},
and are given by $\alpha = - 0.08$ meV $/R$, $\beta = -0.31$ meV $\cos(3\eta)/R$,
and $\hbar v_F \Delta k^{cv}_t = -5.4$ meV $\tau \cos(3\eta)/R^2$
with $R$ the CNT radius in nm, $\tau = K,K'=+,-$, and $\eta$ the chiral angle, $\tan(\eta) = \sqrt{3} N_2/ (2N_1+N_2)$,
for a CNT with chiralities $(N_1,N_2)$.
For $(N_1,N_2)=(20,0)$ we have $R=0.78$ nm, $\alpha = -0.10$ meV, and $\beta = -0.40$ meV.
The induced superconducting gap is $\Delta=0.1$ meV, and the doping of the central
region $-243$ meV. All further parameters are as described in Ref. \onlinecite{Sburset:2011}.
For $(N_1,N_2)=(18,10)$ as used for Fig. 2 in the main text, we have
$R = 0.96$ nm, $\alpha = -0.08$ meV, and $\beta = -0.15$ meV.

\begin{figure}
\begin{center}
	\includegraphics[width=\columnwidth]{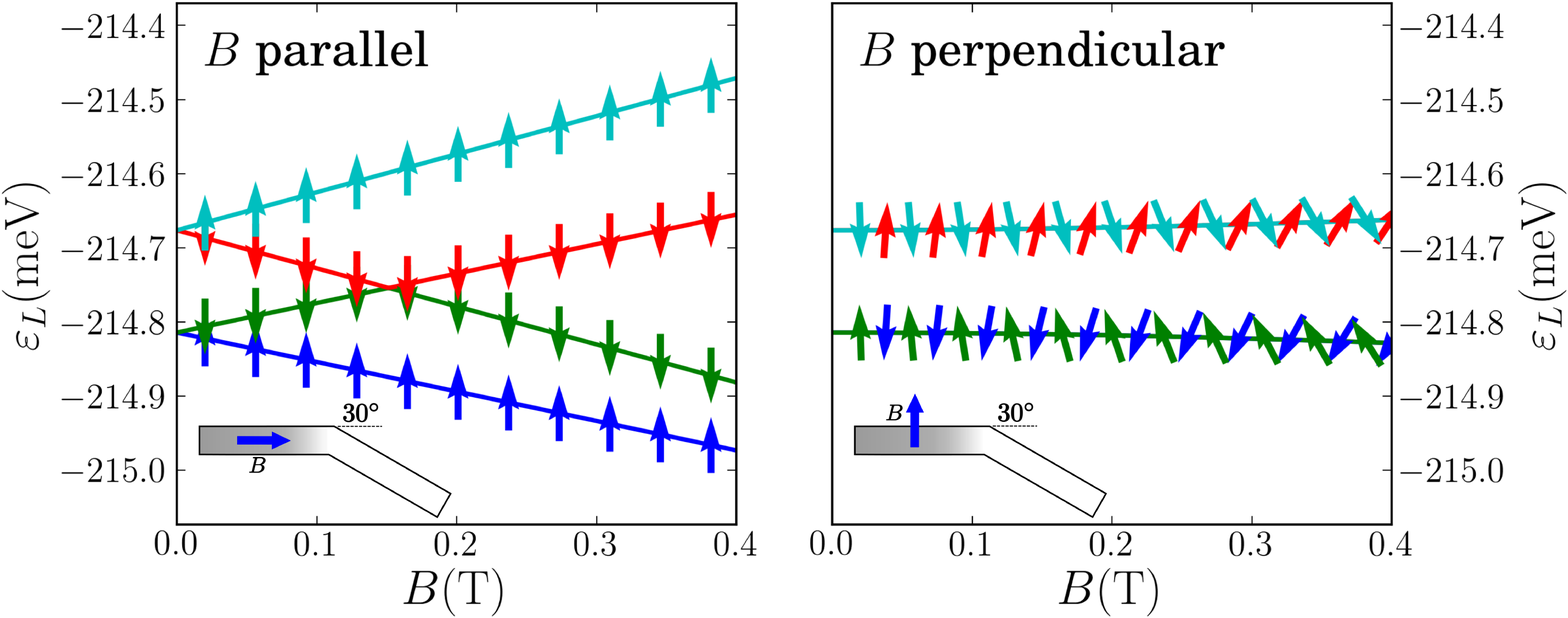}
	\caption{\label{fig:B_1}
	Levels and spin projections of the left QD as a function of parallel and perpendicular magnetic
	fields $B$ for the (18,10) CNT described in Fig. 2 of the main text.
	The sketches in the lower left corners indicate the $B$ field directions with respect to the left QD.
	The arrows indicate the spin projections in the $(x,z)$ plane with the $S^z$ direction pointing
	upwards and the $S^x$ direction to the right in the plots.
	At $B=0$ the levels of
	both valleys are degenerate. At increasing $B$, the levels of one valley increase
	and the levels of the other valley decrease in energy by the combined effect of orbital and
	Zeeman fields. At parallel field, the spins remain parallel to the CNT axis.
	At perpendicular field, the level
	energies are only weakly affected by $B$, yet the spin projections in each valley strongly
	rotate.
	The situations at $B = 0.4$ T correspond to the selected angles $\theta = 0^\circ{}, 90^\circ{}$ in
	the upper right panel of Fig. 2 in the main text.
	}
\end{center}
\end{figure}
\begin{figure}
\begin{center}
	\includegraphics[width=\columnwidth]{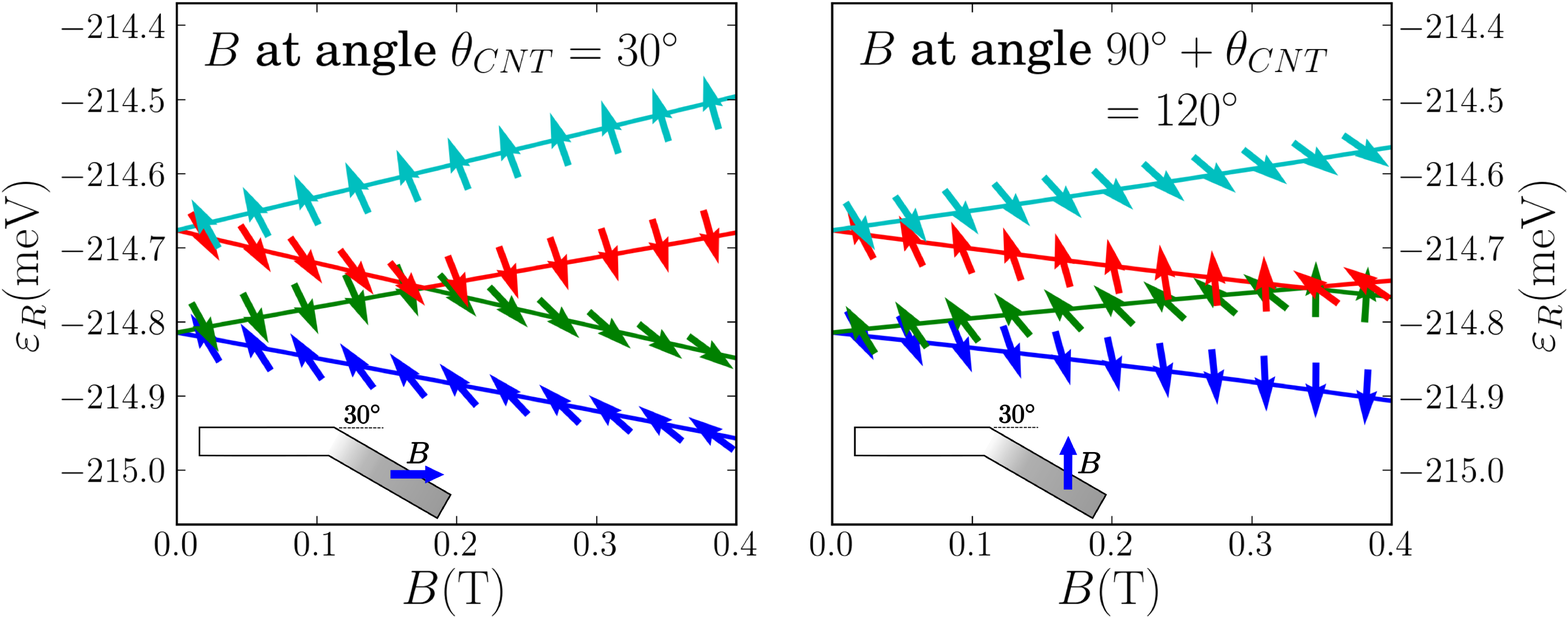}
	\caption{\label{fig:B_2}
	Levels and spin projections as in Fig. \ref{fig:B_1} for the right QD at the angle
	$\theta_{CNT}=30^\circ$ to the left QD. 
	The sketches in the lower left corners indicate the $B$ field directions with
	respect to the right QD. 
	The spins are shown in the global $(x,z)$
	basis corresponding to Fig. \ref{fig:B_1}.
	The situations at $B = 0.4$ T correspond to the selected angles $\theta = 0^\circ{}, 90^\circ{}$ in
	the lower right panel of Fig. 2 in the main text.
	}
\end{center}
\end{figure}
Finally, we illustrate the evolution of the QD levels and their spin polarizations as a function of
the magnetic field $B$. Figure \ref{fig:B_1} displays the 4 spin polarized QD levels of the
(18,10) CNT model used for Fig. 2 in the main text, for magnetic fields parallel and perpendicular
to the CNT axis of the left QD, respectively.
Figure \ref{fig:B_2} shows the levels of the right QD for the same
fields, which are seen for this QD under the additional angle $\theta_{CNT} = 30^\circ$.




\end{document}